\documentclass[twocolumn]{revtex4}
\usepackage{amsmath}
\usepackage{graphicx}
\usepackage{color}
\usepackage{amssymb}
\usepackage{wasysym}
\usepackage{epstopdf}
\usepackage[T1]{fontenc}
\usepackage[utf8]{inputenc}
\usepackage{lmodern}
\usepackage[english]{babel}
\usepackage{bm}% bold math
\usepackage[final]{pdfpages}
\input{epsf}

\begin{document}

\title{Biaxial Nematics of Hard Cuboids in an External Field}
\author{Alejandro Cuetos$^1$, Effran Mirzad Rafael$^2$, Daniel Corbett$^2$, and Alessandro Patti$^{2,}$}
\email{alessandro.patti@manchester.ac.uk}
\affiliation{$^1$Department of Physical, Chemical and Natural Systems, Pablo de Olavide University, 41013 Sevilla, Spain}
\affiliation{$^2$School of Chemical Engineering and Analytical Science, The University of Manchester, Manchester, M13 9PL, UK}

\begin{abstract}
We investigate the phase behavior of colloidal suspensions of board-like particles under the effect of an external field and assess the still disputed occurrence of the biaxial nematic (N$_\text{B}$) liquid crystal phase.  The external field promotes the rearrangement of the initial isotropic (I) or uniaxial nematic (N$_\text{U}$) phase and the formation of the N$_\text{B}$ phase. In particular, very weak field strengths are sufficient to spark a direct I-N$_\text{B}$ or N$_\text{U}$-N$_\text{B}$ phase transition at the self-dual shape, where prolate and oblate particle geometries fuse into one. By contrast, forming the N$_\text{B}$ phase at any other geometry requires stronger fields and thus reduces the energy efficiency of the phase transformation. Our simulation results show that self-dual shaped board-like particles with moderate anisotropy are able to form N$_\text{B}$ liquid crystals under the effect of a surprisingly weak external stimulus and suggest a path to exploit low-energy uniaxial-to-biaxial order switching.

\end{abstract}

\maketitle

It is well established that anisotropic colloidal particles can self-assemble into a plethora of fascinating liquid crystal (LC) phases in a solvent. Onsager showed that mere excluded volume effects can force hard-core particles to align along a common director at sufficiently large concentrations \cite{onsager}. The resulting LC phases found at the thermodynamic equilibrium and their structural properties strongly depend on the particle geometry. In particular, prolate particles tend to orient along their major axis, while oblate particles along their minor axis. Although this tendency is regularly observed in systems of uniaxial particles, such as disks, whose orientation is determined by a single unit vector, it is less predictable for biaxial particles, such as cuboids, whose orientation is fully determined by two unit vectors. For instance, slightly oblate hard board-like particles (HBPs) have been shown to orient along their major axis and thus form prolate (rather than oblate) smectic LC phases \cite{cuetos}. This unusual arrangement was observed in suspensions of HBPs with length-to-thickness ratio $L^*\equiv L/T=12$ and width-to-thickness ratio  $W^* \equiv W/T \approx \sqrt{L^*}$, a geometry where oblate and prolate shapes fuse into one.

Such an exclusive particle architecture, referred to as self-dual shape, was predicted to favour the formation of the biaxial nematic (N$_\text{B}$) phase in systems of HBPs with rounded \cite{taylor} or square \cite{belli} corners. Nevertheless, these theoretical predictions were made within the context of the Zwanzig model, which does not allow free rotations of particles and restricts their orientations to only six. Computer simulations that explored the phase behavior of freely rotating HBPs highlighted the challenge of observing stable N$_\text{B}$ phases, even when an element of size-dispersity is incorporated \cite{patti1}. The very recent and insightful simulation study by the Dijkstra's group showed that monodisperse HBPs, with a geometry close or equal to the self-dual shape, are able to stabilise the N$_\text{B}$ phase only if their anisotropy is significantly large, with $L^*\ge23$ \cite{dussi}. Nevertheless, the highly uniform colloidal board-like particles synthesised by Nie and coworkers were not observed to assemble into the N$_\text{B}$ phase in a wide spectrum of aspect ratios, between $L^*=20$ and 180, and very close to the self-dual shape \cite{nie}. A direct observation of a stable N$_\text{B}$ phase was reported almost ten years ago in dispersions of purely repulsive and quasi dual-shaped goethite particles \cite{vroege}, whose stability was shown to depend on the extremely large particle size dispersity \cite{belli}. These computational and experimental findings unambiguously indicate that self-dual shaped HBPs cannot self-assemble into an N$_\text{B}$ phase, unless their anisotropy and/or size dispersity are especially relevant. Vroege and coworkers investigated the effect of a magnetic field on the phase behavior of quasi-dual-shaped polydisperse goethite particles in suspension and observed a biaxial-to-uniaxial nematic transition above a certain magnetic field strength \cite{leferink}. To explain the origin of these experimental results, the same authors formulated a mean-field theory to calculate the phase diagram of HBPs in the presence of a magnetic field. Despite the good qualitative agreement between theory and experiments, the former neglects the effect of size polydispersity, being crucial to stabilise the N$_\text{B}$ phase \cite{vroege}, and is limited by the strong approximations imposed by the Zwanzig model, which was shown to be not especially accurate to describe the phase behavior of HBPs \cite{cuetos, patti1}.

In this work, we perform Monte Carlo (MC) simulations of freely-rotating monodisperse HBPs in the presence of an external field that promotes a phase transition from the isotropic (I) or uniaxial nematic (N$_\text{U}$) phase to the N$_\text{B}$ phase. Because our main interest is gaining an insight into the energetics associated to this process, the field strength is a simulation parameter that assumes values between $\epsilon_f^* \equiv \epsilon_f \beta=0.1$ and 3, with $\beta$ the inverse temperature. At the same time, we propose an alternative route for the formation of the N$_\text{B}$ phase that does not necessarily imply extreme anisotropies \cite{dussi} or particle size dispersities \cite{vroege}, nor the addition of depletants \cite{belli2}. In the following, we discuss the main aspects of the model and simulation methodology applied. The interested reader is referred to our previous works for additional details \cite{cuetos, patti1}.

We have simulated fluids of HBPs with $L^*=12$ and $1 \le W^* \le 12$, including $W^*=\sqrt{L^*} \approx 3.46$, which gives the above mentioned self-dual shape. The particle thickness, $T$, is our system unit length and is kept constant. The unit vectors, $\hat{\textbf{x}}$, $\hat{\textbf{y}}$, and $\hat{\textbf{z}}$, aligned along $T$, $W$, and $L$, respectively, define the particle orientation in space. Interactions are described by a hard-core potential, which only depends on the distance between pairs of particles. Consequently, an MC move is rejected if an overlap occurs and accepted otherwise. The separating axes method, introduced by Gottschalk \textit{et al.} \cite{gottschalk} and later modified to study suspensions of tetragonal parallelepipeds \cite{john}, has been employed to assess the occurrence of overlaps between HBPs. The interaction between HBPs and external field is described by a potential energy term that favours the alignment of the particle intermediate axis, $\hat{\textbf{y}}$, with the direction of field, indicated by $\hat{\textbf{e}}$, and reads

\begin{equation}\label{eq1}
U_{\text{ext}} = \frac{\epsilon_f}{2} \sum_{i=1}^N (1-3\cdot(\hat{\textbf{y}_i}\cdot\hat{\textbf{e}})^2)
\end{equation}

\noindent where $N$ indicates the number of HBPs in the system, which is in the range [1100, 3648], depending on $W^*$. The field direction, $\hat{\textbf{e}}$, is always aligned with the same box axis, and the field itself coupled to the particle axis $\hat{\textbf{y}}$ in all the systems studied. The latter could have been alternatively coupled to $\hat{\textbf{x}}$ or $\hat{\textbf{z}}$, but this choice would have promoted the formation of, respectively, oblate (N$^-_\text{U}$) or prolate (N$^+_\text{U}$) uniaxial nematics, which are normally observed without the application of an external field \cite{cuetos}. We stress that there is no intention here to mimic a real electric field, whose effect on a colloidal suspension depends on particle polarizability, charge and dielectric constant difference with that of the solvent, nor a magnetic field, implying suitable particle magnetic susceptibility and dipole moment. Rather than on the physical nature of the external stimulus, which we keep as simple as possible, we focus on its resulting effect, that is the ability of reorienting and aligning HBPs as well as the formation of uniaxial and biaxial LC phases. How electric and magnetic fields can be employed to control the internal organisation and self-assembly of colloidal suspensions has been explored in recent theoretical and simulation studies on spherical \cite{hynninen1, hynninen2, smallenburg1, smallenburg2} and anisotropic \cite{kwaadgras1, kwaadgras2, troppenz, azari} particles, including HBPs within the Zwanzig model approximation \cite{leferink}. The field-particle potential in Eq.\ (1) can be regarded as an effective potential that promotes the alignment of the particle intermediate axis with $\hat{\textbf{e}}$. Finally, we assume the inter-particle interactions to be predominantly governed by excluded-volume effects, ignoring any enthalpic contribution that could arise from the presence of the field. 

The external field has been applied to suspensions of HBPs that were either in dense I phases, very close to the border with the nematics, or deeply in the N$^-_\text{U}$ or N$^+_\text{U}$ phases, at the packing fraction $\eta \equiv Nv/V=0.34$, with $v$ and $V$ the particle and box volume, respectively. At this packing fraction, the nematic phase is stable for the whole range of particle geometries studied here. In particular, N$^+_\text{U}$ and N$^-_\text{U}$ phases are observed, respectively, at $1 \le W^* < 3.46$ and $3.46 < W^* \le 12$, as shown in the phase diagram available in Ref.\ \cite{cuetos}. By contrast, the packing fraction of the systems in the I phase changes with $W^*$, between $\eta=0.20$ and 0.31. All simulations have been run in the $NVT$ ensemble, where number of particles, volume and temperature are kept constant. After equilibrating a number of I and N$_\text{U}$ phases, the field was switched on and new equilibrium states were achieved.

To identify the long-range order of the phases at the initial and final equilibrium states, we calculated the nematic order parameter and nematic director coupled to each of the three particle axes. In particular, we applied the standard procedure of diagonalising the following traceless symmetric second-rank tensor \cite{eppenga}

\begin{equation}
\label{eq2}
{\bf Q}^{\lambda \lambda}= \frac{1}{2N} \left\langle \, \sum^{N}_{i=1} (3{{\hat{\lambda}}}_{i} \cdot {{\hat{\lambda}}}_{i}
\,-\,{\bf I})\, \right\rangle
\end{equation}

\noindent where $i$ indicates a generic HBP, ${{\hat{\lambda}}}$=$\bf{{\hat{x}}}$, $\bf{{\hat{y}}}$, $\bf{{\hat{z}}}$ is the particle unit orientation vector and ${\bf I}$ the second-rank unit tensor. The resulting eigenvalues ($S_{2,W}$, $S_{2,T}$, $S_{2,L}$) and associated eigenvectors ($\bf{\hat{m}}$, $\bf{\hat{p}}$, $\bf{\hat{n}}$) identify, respectively, the order parameters and nematic directors coupled to each particle axis. If the order is very weak, as in the I phase, the eigenvalues vanish and the eigenvectors are meaningless. However, if a preferential direction of alignment exists, this is identified by the largest positive eigenvalue of ${\bf Q}^{\lambda \lambda}$ and its corresponding eigenvector. For instance, the N$^+_\text{U}$ phase is characterised by a large value of $S_{2,L}$ and a preferential alignment of the particle $\bf{\hat{z}}$ axes along the director $\bf{\hat{n}}$. The three tensors in Eq.\ (2) can be applied to estimate three additional order parameters, $B_{2,W}$, $B_{2,T}$ and $B_{2,L}$, coupled, respectively, to $\bf{\hat{m}}$, $\bf{\hat{p}}$ and $\bf{\hat{n}}$, that identify the occurrence of biaxiality \cite{camp}. However, it is sufficient to monitor the fluctuations of the unit vectors that are perpendicular to the nematic director associated to the largest eigenvalue of ${\bf Q}^{\lambda \lambda}$ \cite{teixeira, allen, camp2}. If $S_{2,L}$ is the largest eigenvalue, then these fluctuations are quantified by the biaxial order parameter  $B_{2,L}=(\hat{{\bf p}}\cdot{\bf Q}^{xx}\cdot \hat{{\bf p}}+\hat{{\bf m}}\cdot{\bf Q}^{yy} \cdot \hat{{\bf m}}-\hat{{\bf p}} \cdot {\bf Q}^{yy} \cdot \hat{{\bf p}} - \hat{{\bf m}} \cdot {\bf Q}^{xx} \cdot \hat{{\bf m}})/3$. In this case, $S_{2,L}=1$ and $B_{2,L}=0$ would indicate perfect alignment along $\bf{\hat{n}}$ only and the occurrence of the N$^+_\text{U}$ phase, while $S_{2,L}=1$ and $B_{2,L}=1$ would indicate perfect alignment along three directors and the formation of the N$_\text{B}$ phase. Finally, to unambiguously discard the occurrence of smectic LC phases, we have calculated the pair distribution functions along the relevant nematic directors, which do not show any significant periodicity as expected from the relatively low packing.

In the light of these preliminary considerations, we now discuss our simulation results, which depend on three main factors: (\textit{i}) the initial state of the system, being either isotropic or uniaxial nematic, (\textit{ii}) the field strength, $\epsilon_f$, and (\textit{iii}) the particle geometry, which only depends on $W^*$. In Figs.\ 1 and 2, we show the phases observed, in the new equilibrium state, when a field of strength $0.1 \le \epsilon_f^* \le 3$ is applied to, respectively, I or N$_\text{U}$ phases of HBPs with $1 \le W^* \le 12$. The values of the uniaxial  and biaxial nematic order parameters that validate these results are provided in the Supporting Information. The state points at $\epsilon_f=0$ are included as a reference and identify I phases in Fig.\ 1, and  N$_\text{U}^+$ ($W^*< 3.46$) or N$_\text{U}^-$ ($W^*>3.46$) phases in Fig.\ 2. The criterion adopted in both figures to classify the new equilibrium phases essentially depends on the value of the uniaxial and biaxial order parameters and is summarised in Table I. 

\begin{figure}
\centering
  \includegraphics[width=1.0\columnwidth]{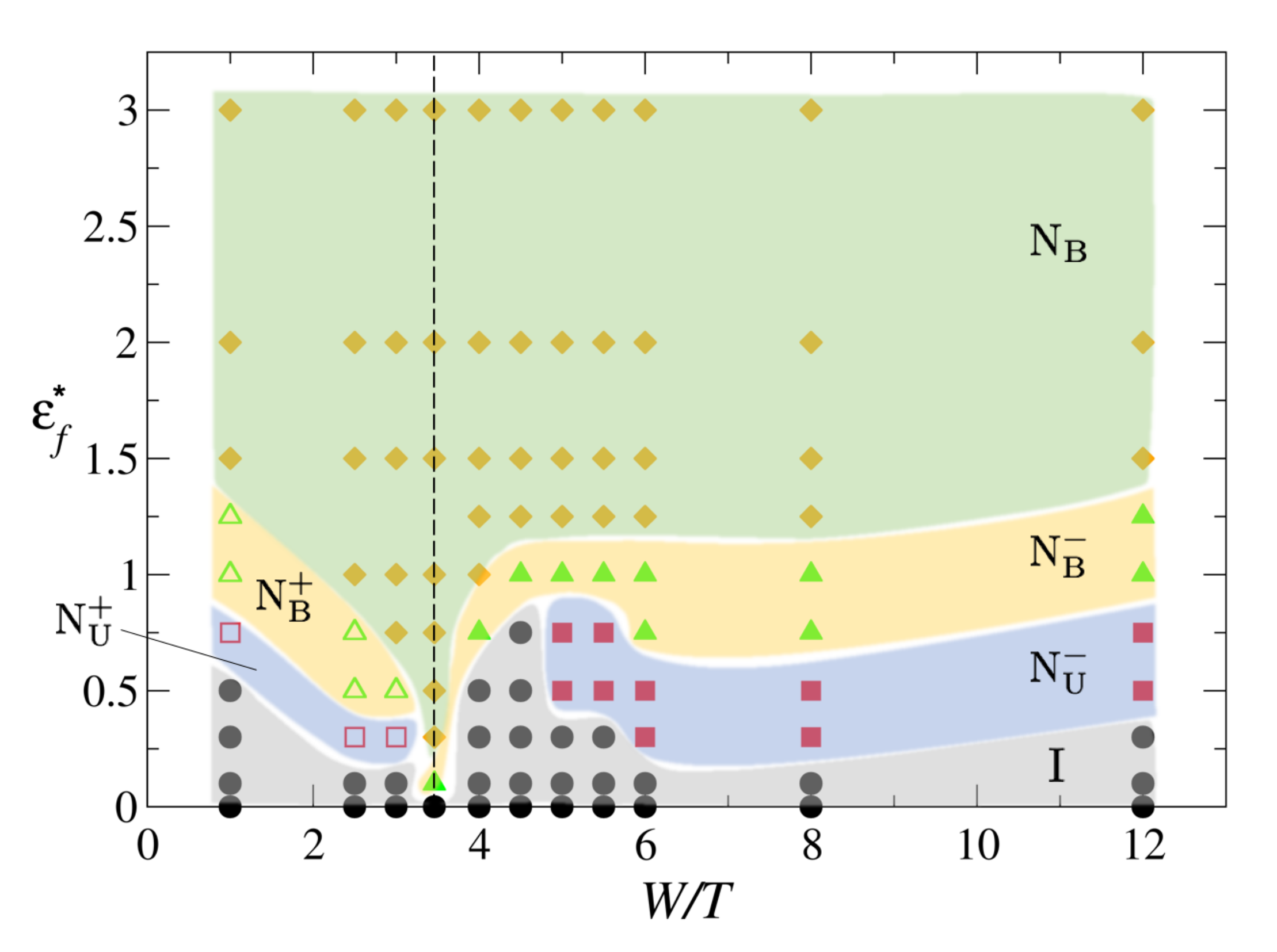}
  \caption{Equilibrium LCs resulting from the application of the external field given in Eq.\ (1), with strength $\epsilon_f$, to an isotropic phase of HBPs of length-to-thickness ratio $L^*=12$ and width-to-thickness ratio $1 \le W^* \le 12$. Circles, empty and solid squares, empty and solid triangles, and diamonds indicate, respectively, I, N$^+_\text{U}$, N$^-_\text{U}$, N$^+_\text{B}$, N$^-_\text{B}$, and N$_\text{B}$ phases. Shaded areas are guides for the eye. The vertical dashed line refers to the self-dual shape, where $W=\sqrt{LT}$.}
  \label{fgr:fig1}
\end{figure}

\begin{figure}
\centering
  \includegraphics[width=1.0\columnwidth]{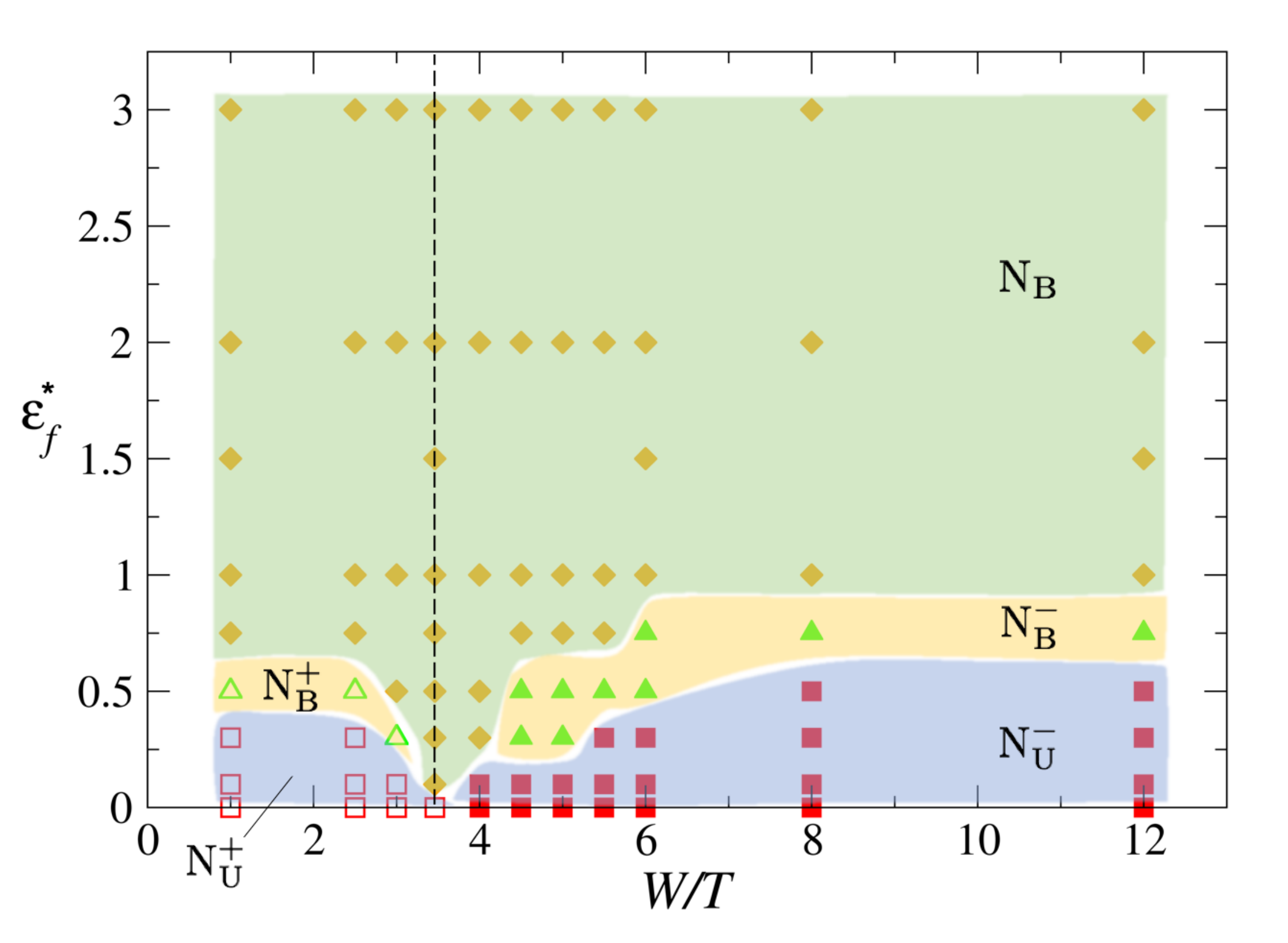}
  \caption{Equilibrium LCs resulting from the application of the external field given in Eq.\ (1), with strength $\epsilon_f$, to uniaxial nematic phases of HBPs of length-to-thickness ratio $L^*=12$ and width-to-thickness ratio $1 \le W^* \le 12$. Circles, empty and solid squares, empty and solid triangles, and diamonds indicate, respectively, I, N$^+_\text{U}$, N$^-_\text{U}$, N$^+_\text{B}$, N$^-_\text{B}$, and N$_\text{B}$ phases. Shaded areas are guides for the eye. The vertical dashed line refers to the self-dual shape, where $W=\sqrt{LT}$.}
  \label{fgr:fig1}
\end{figure}

According to this criterion, two \textit{weak} biaxial phases, similar to those previously found by the Patras group in suspensions of spheroplatelets \cite{peroudikis1, peroudikis2, peroudikis3}, have been identified. To distinguish \textit{weak} and \textit{strong} biaxial phases, we look at the magnitude of the biaxial order parameter that is associated to the largest uniaxial order parameter. If the latter is, for example, $S_{2,L}$, then the former is $B_{2,L}$. In a strong biaxial phase, simply labelled as N$_\text{B}$, the HBPs are highly orientationally ordered, with their three unit vectors $\bf{{\hat{x}}}$, $\bf{{\hat{y}}}$ and $\bf{{\hat{z}}}$ oriented along three mutually perpendicular directions. This significant degree of orientational order is reflected in the large values of all the uniaxial and biaxial order parameters. By contrast, the weak biaxial phases, labelled as N$^+_\text{B}$ and N$^-_\text{B}$, show a modest, but not negligible biaxial order parameter, between 0.20 and 0.35, and at least two pronounced uniaxial order parameters, above 0.40, which give them a prolate or oblate character, respectively.

\begin{table}[h]
\setlength{\tabcolsep}{3.5pt}
\caption{Range of uniaxial and relevant biaxial order parameters for isotropic (I), uniaxial (N$_\text{U}$) and biaxial (N$_\text{B}$) phases observed in suspensions of HBPs. The superscripts $^+$ and $^-$ indicate, respectively, prolate and oblate phase symmetry. (\textit{a}) At least one between $S_{2,T}$ and $S_{2,W}$ should be in the range specified. (\textit{b}) At least one between $S_{2,L}$ and $S_{2,W}$ should be in the range specified.}
{\begin{tabular}  {@{}c|c|c|c|c|c}
\toprule
$S_{2,L}$  & $S_{2,T}$  & $S_{2,W}$ & $B_{2,L}$ $\vee$ $B_{2,T}$ & Phase &\\
\colrule
0.00 - 0.20 & 0.00 - 0.20 & 0.00 - 0.20 & -- & I   \\
0.40 - 1.00 & 0.00 - 0.35 & 0.00 - 0.35 & 0.00 - 0.30 & N$^+_\text{U}$ \\
0.00 - 0.35 & 0.40 - 1.00 & 0.00 - 0.35 & 0.00 - 0.30 & N$^-_\text{U}$ \\
0.40 - 1.00 & 0.35 - 1.00 & 0.35 - 1.00 & 0.20 - 0.35 & N$^+_\text{B}$ & {\footnotesize \textit{a}}\\
0.35 - 1.00 & 0.40 - 1.00 & 0.35 - 1.00 & 0.20 - 0.35 & N$^-_\text{B}$ & {\footnotesize \textit{b}}\\
0.40 - 1.00 & 0.40 - 1.00 & 0.40 - 1.00 & 0.35 - 1.00 & N$_\text{B}$ \\
\botrule
\end{tabular}}
\end{table}

An external field with strength $\epsilon_f^* < 1$ is able to reorient prolate and oblate HBPs along a common direction to form N$^+_\text{U}$ and N$^-_\text{U}$ phases, respectively (see Fig.\ 1). Interestingly enough, at the self-dual shape, even a very gentle field is sufficient to spark a direct I-N$^-_\text{B}$ transition and then, at $\epsilon_f^* = 0.3$, the formation of the N$_\text{B}$ phase, which becomes dominant at larger field strengths across the complete set of particle geometries. Similar tendencies are noticed when the same field is applied to a uniaxial nematic phase. In this case, however, the I-N$_\text{B}$ phase transition is detected at an almost insignificant field strength, $\epsilon_f^*=0.1$, with no intermediate formation of weak biaxial phases (see Fig.\ 2). Snapshots showing the phase order before and after the application of a relatively weak field to I and N$_\text{U}$ phases are shown in Fig.\ 3.  

\begin{figure}
\centering
  \includegraphics[width=1.0\columnwidth]{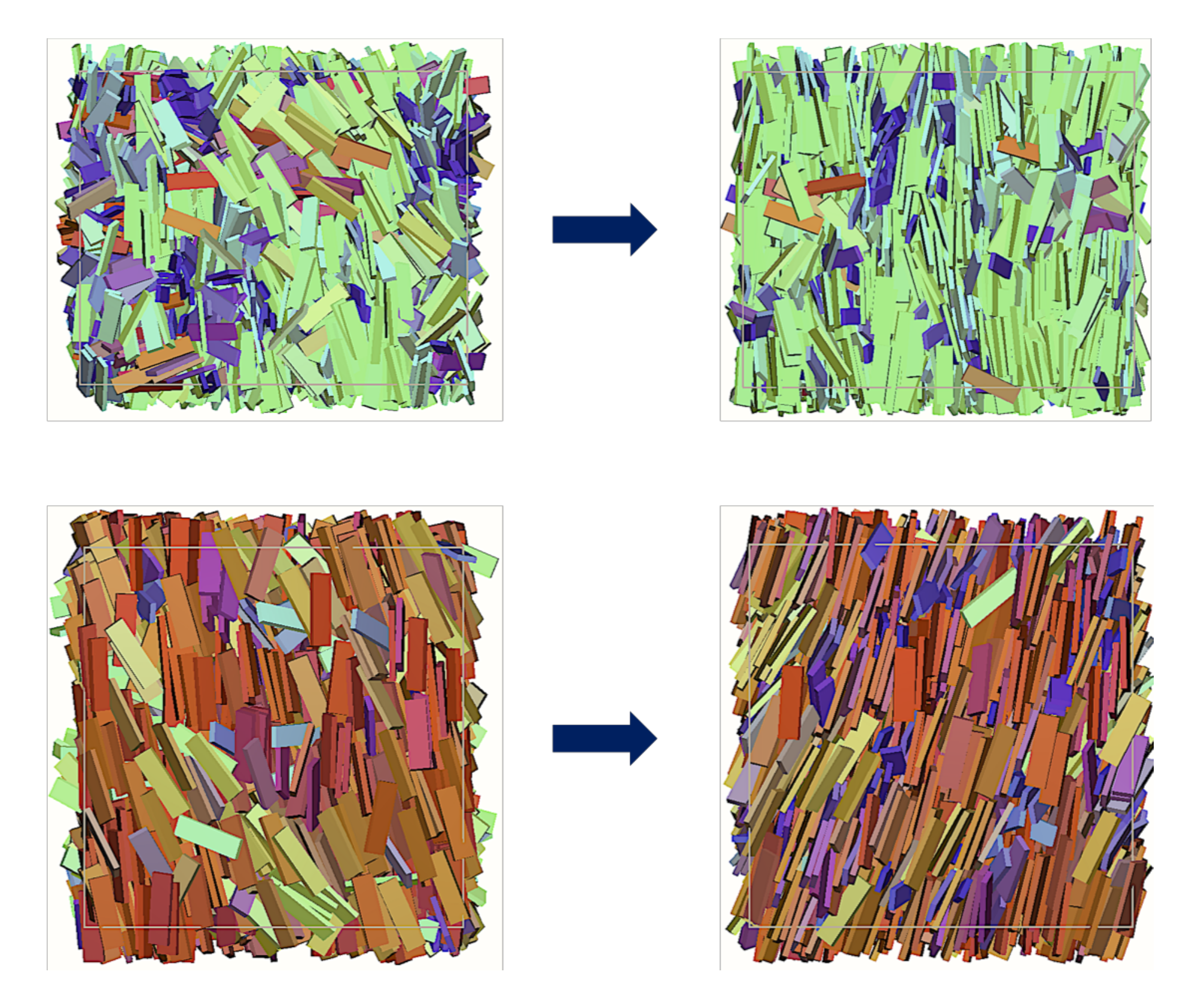}
  \caption{Direct I-to-N$_\text{B}$ (top) and N$^-_\text{U}$-to-N$_\text{B}$ (bottom) phase transitions induced by an external field with strength $\epsilon^*_f=0.3$. Particles have self-dual shape with relative length $L^*=12$ and relative width $W^*=\sqrt{L^*} \approx 3.46$. The packing fraction of the isotropic and uniaxial nematic phase is, respectively, $\eta=0.31$ and 0.34. The colour gradient indicates the orientation of the particle major axis.}
  \label{fgr:fig3}
\end{figure}

Therefore, regardless of the initial system configuration, the dual-shape is found to be the optimal particle geometry to form the N$_\text{B}$ phase in terms of energy costs as the strength of the field applied at $W^*=\sqrt{L^*}$ is at its minimum. In spite of being relatively low, such energy costs are crucial to achieve the desired orientational order with no increase of the translational order, which remains negligible and would otherwise imply the onset of biaxial smectic, rather than nematic, LCs. As a matter of fact, the calculation of the pair distribution functions along the nematic director, $g_{\parallel}(r)$, excludes the formation of positionally ordered phases (see Supporting Information). At particle geometries that are different, but still relatively similar, to the self-dual shape, stabilising the N$_\text{B}$ phase becomes more energetically demanding, whether the initial phase is I or N$_\text{U}$. For instance, an I phase incorporating HBPs with $W^*=4$ requires a field strength of approximately $\epsilon_f^*=1$. At larger anisotropies, stronger fields are required, up to $\epsilon_f^*=1.5$ at $W^*=12$. As expected, the energy to reorient HBPs originally in uniaxial nematics into biaxial nematics is sensibly lower. At $W^*=4$, this is approximately 50\% of that needed to align them in the isotropic phase.

In summary, our results suggest that it is indeed possible to observe biaxial nematics in colloidal suspensions of monodisperse HBPs. To this end, one needs to apply an external field whose strength is comparable to the particle thermal energy. In particular, the self-dual shape is the most appropriate particle geometry for energy-efficient I-N$_\text{B}$ and N$_\text{U}$-N$_\text{B}$ phase transitions. Although these conclusions agree well with past theoretical works \cite{straley, mulder, taylor, belli}, \textit{de facto} they indicate that this agreement is the consequence of restricting particle orientations and/or neglecting the occurrence of positionally ordered phases, which artificially enhance the stability of the N$_\text{B}$ phase. Mere excluded volume effects were shown to be insufficient to observe biaxial nematics in monodisperse systems of HBPs \cite{cuetos}, unless extremely long particles are employed \cite{dussi}. Applying an external field to monodisperse freely-rotating HBPs produces a scenario that the above mentioned theories, despite their strong assumptions, had predicted: the N$_\text{B}$ phase can be stabilised at moderate particle anisotropies and a direct I-N$_\text{B}$ transition, so far only detected experimentally in systems of significantly size dispersed goethite particles \cite{vroege}, observed. The external field induces the desired alignment of particles with no effect on their positional order an at a relatively low energy cost. While promoting the particle alignment along one direction, this field does not prevent them from rotating freely, especially around the remaining two directions. By gradually increasing the field strength, the particles, regardless their geometry, assume a more and more narrow distribution of orientations, which eventually end up fluctuating around three main orthogonal directions, being the signature of the formation of the N$_\text{B}$ phase.

AC acknowledges project P12-FQM-2310 funded by the Junta de Andaluc\'{\i}a-FEDER and C3UPO for HPC facilities provided. AC also acknowledges grant PPI1719 awarded by the Pablo de Olavide University for funding his research visit to the School of Chemical Engineering and Analytical Science, The  University of  Manchester. EMR acknowledges the Malaysian Government Agency Majlis Amanah Rakyat for funding his PhD at the University of Manchester. DC and AP acknowledge financial support from EPSRC under grant agreement EP/N02690X/1.

\end{document}